\documentclass[a4paper,fleqn,usenatbib]{mnras}

\usepackage{mathptmx}

\usepackage[T1]{fontenc}
\usepackage{ae,aecompl}

\usepackage{graphicx}	
\usepackage{amsmath}	
\usepackage{amssymb}	

\title[The pulsations of G~207-9 and LP~133-144]{G~207-9 and LP~133-144: light curve analysis and asteroseismology of two ZZ Ceti 
stars
}

\author[Zs. Bogn\'ar et al.]{
Zs. Bogn\'ar,$^{1}$\thanks{E-mail: bognar.zsofia@csfk.mta.hu (Zs.B.)}
M. Papar\'o,$^{1}$
L. Moln\'ar,$^{1}$
P.~I. P\'apics,$^{2}$
E. Plachy,$^{1}$
E. Vereb\'elyi,$^{1}$
\newauthor and \'A. S\'odor$^{1}$
\\
$^{1}$Konkoly Observatory, MTA Research Centre for Astronomy and Earth Sciences, Konkoly Thege Mikl\'os \'ut 15-17, H--1121 Budapest\\
$^{2}$Instituut voor Sterrenkunde, KU Leuven, Celestijnenlaan 200D, B-3001 Leuven, Belgium
}

\date{Accepted XXX. Received YYY; in original form ZZZ}

\pubyear{2016}

\begin{document}
\label{firstpage}
\pagerange{\pageref{firstpage}--\pageref{lastpage}}
\maketitle

\begin{abstract}
G~207-9 and LP~133-144 are two rarely observed ZZ Ceti stars located in the middle and close to the blue edge of 
the ZZ Ceti instability domain, respectively. We aimed to observe them at least during one observing season at
Konkoly Observatory with the purpose of extending the list of known pulsation modes for asteroseismic investigations
and detect any significant changes in their pulsational behaviour. We determined five and three new 
normal modes of G~207-9 and LP~133-144, respectively. In LP~133-144, our frequency analysis also revealed
that at least at three modes there are actually triplets with frequency separations of $\sim4\mu$Hz. 
The rotational period of LP~133-144 based on the triplets is $\simeq42$\,h.
The preliminary asteroseismic
fits of G~207-9 predict $T_{\rmn{eff}}=12\,000$ or $12\,400$\,K and $M_*=0.855-0.870\,M_{\sun}$ values for the effective
temperature and mass of the star, depending on the assumptions on the spherical degree ($l$) values of the modes. 
These results are in agreement with the spectroscopic determinations. In the case of LP~133-144,
the best-fitting models prefer $T_{\rmn{eff}}=11\,800$\,K in effective temperature and
$M_*\geq0.71\,M_{\sun}$ stellar masses, which are more than $0.1\,M_{\sun}$ larger than the spectroscopic value.
\end{abstract}




\section{Introduction}


ZZ Ceti (or DAV) stars constitute the most populated group of pulsating white dwarfs.
Their light variations are the results of local changes in their surface temperatures due to the excitation of 
nonradial $g$-mode pulsations in their non-degenerate envelope. This envelope
consists of an inner helium and an outer hydrogen layer, therefore, the hydrogen Balmer-lines dominate
the spectra of ZZ Ceti stars. The pulsations are driven by the so-called `convective driving' mechanism 
\citep{1991MNRAS.251..673B, 1999ApJ...511..904G}, as the driving region is associated with the base of the envelope convection zone.
Otherwise, pulsating white dwarfs are just like their non-pulsating counterparts, and the information we gain on white 
dwarf structures by asteroseismic investigations can be essential to understand white dwarfs as a whole group.

ZZ Ceti stars are short-period and low-amplitude pulsators with $11\,000-13\,000$\,K effective temperatures and 
modes excited typically in the 
$100-1400$\,s period range with $\sim$mmag amplitudes. However, within this period and amplitude range the stars
exhibit a large variety of pulsational behaviour, from the star showing one rotational triplet only (G~226-29, \citealt{1995ApJ...447..874K}) 
to the `rich' DA pulsators with more than a dozen normal modes known (see e.g. \citealt{2009AIPC.1170..621B}). 
Temporal variations in their pulsational behaviour are also well documented
in white dwarfs, e.g. the case of GD~154 showing once a strongly non-sinusoidal light curve with one dominant mode 
and a series of its harmonic and near-subharmonic ($\sim n/2f_i$) peaks in its Fourier transform, or just behaving
as a simple multiperiodic pulsator another time \citep{1978ApJ...220..614R, 2013MNRAS.432..598P}. For comprehensive reviews of 
the observational and theoretical aspects of pulsating white dwarf studies, see the papers of \citet{2008PASP..120.1043F, 2008ARA&A..46..157W}
and \citet{2010A&ARv..18..471A}.
We also refer to the work of \citet{2013ApJ...762...57V}, in which the authors successfully reconstructed the
boundaries of the empirical ZZ Ceti instability strip applying theoretical calculations, including its extension
to lower effective temperatures and surface gravities, that is, further to the domain of the 
extremely low-mass DA pulsators.

White dwarf observations with the \textit{Kepler} space telescope revealed another new feature in 
ZZ Ceti stars, namely recurring increases in the stellar flux (`outbursts') in two cool DAVs
being close to the red edge of the instability strip \citep{2015ApJ...809...14B, 2015ApJ...810L...5H}.

G~207-9 and LP~133-144 were observed as part of our project aiming at least one-season-long local photometric 
time series measurements of white dwarf pulsators. Our purposes are to examine the short-term variability of pulsation modes 
in amplitude and phase, and to obtain precise periods for asteroseismic investigations. 
We have already published our findings on two cool
ZZ Ceti stars (KUV~02464+3239, \citealt{2009MNRAS.399.1954B} and GD~154, \citealt{2013MNRAS.432..598P}), one
ZZ Ceti located in the middle of the instability strip (GD~244; \citealt{2015ASPC..493..245B}), and
the DBV type KUV~05134+2605 \citep{2014A&A...570A.116B}. With the observations of G~207-9 and LP~133-144,
we extended our scope of investigations to higher effective temperatures in the DAV instability domain. 

\section{Observations and data reduction}

We collected photometric data both on G~207-9 ($B=14.8$\,mag, 
$\alpha_{2000}=18^{\mathrm h}57^{\mathrm m}30^{\mathrm s}$, $\delta_{2000}=+33^{\mathrm d}57^{\mathrm m}25^{\mathrm s}$) 
and LP~133-144 ($B=15.5$\,mag, 
$\alpha_{2000}=13^{\mathrm h}51^{\mathrm m}20^{\mathrm s}$, $\delta_{2000}=+54^{\mathrm d}57^{\mathrm m}43^{\mathrm s}$)
in the 2007 observing season.
We used the 1-m Ritchey-Chr\'etien-Coud\'e telescope at Piszk\'estet\H o mountain station of
Konkoly Observatory. The detector was a Princeton Instruments 
VersArray:1300B back-illuminated CCD camera. The measurements were made in white light and with
10 or 30\,s integration times, depending on the weather conditions.

We observed G~207-9 and LP~133-144 on 24 and 28 nights, respectively. Tables~\ref{table:logg207} and \ref{table:loglp}
show the journals of observations. Altogether, 85 and 137\,h of photometric data were collected on G~207-9 and 
LP~133-144, respectively.

\begin{table}
\centering
\caption{Journal of observations of G~207-9. `Exp.' is the exposure time used.}
\label{table:logg207}
\begin{tabular}{p{3.5mm}rccrr}
\hline
Run & UT date & Start time & Exp. & Points & Length\\
no. & (2007) & (BJD-2\,450\,000) & (s) &  & (h)\\
\hline
01 & Mar 26 & 4185.540 & 30 & 279 & 2.57\\
02 & Mar 27 & 4186.533 & 30 & 280 & 2.70\\
03 & Apr 02 & 4192.527 & 10 & 743 & 2.70\\
04 & Apr 03 & 4193.540 & 30 & 208 & 1.88\\
05 & Jun 15 & 4267.351 & 10 & 1001 & 4.46\\
06 & Jun 16 & 4268.374 & 30 & 427 & 4.10\\
07 & Jun 17 & 4269.396 & 10 & 1028 & 3.68\\
08 & Jun 18 & 4270.341 & 30 & 500 & 4.75\\
09 & Jun 19 & 4271.345 & 10 & 1173 & 4.89\\
10 & Jun 20 & 4272.454 & 30 & 198 & 2.23\\
11 & Jul 06 & 4288.389 & 30 & 184 & 2.42\\
12 & Jul 07 & 4289.347 & 10 & 1417 & 5.18\\
13 & Jul 08 & 4290.337 & 10 & 1449 & 5.60\\
14 & Jul 09 & 4291.351 & 30 & 34 & 0.30\\
15 & Jul 10 & 4292.434 & 30 & 227 & 2.27\\
16 & Jul 26 & 4308.434 & 30 & 349 & 3.12\\
17 & Jul 27 & 4309.337 & 10 & 1071 & 3.69\\
18 & Jul 30 & 4312.452 & 30 & 134 & 1.42\\
19 & Jul 31 & 4313.312 & 30 & 492 & 5.03\\
20 & Aug 01 & 4314.460 & 10 & 885 & 3.05\\
21 & Aug 10 & 4323.327 & 30 & 46 & 0.40\\
22 & Aug 13 & 4326.323 & 30 & 721 & 6.43\\
23 & Aug 14 & 4327.341 & 10 & 1646 & 5.74\\
24 & Aug 15 & 4328.315 & 10 & 1896 & 6.53\\
\multicolumn{2}{l}{Total:} & & \multicolumn{2}{r}{16\,388} & 85.16\\
\hline
\end{tabular}
\end{table}

\begin{table}
\centering
\caption{Journal of observations of LP~133-144. `Exp.' is the exposure time used.}
\label{table:loglp}
\begin{tabular}{p{3.5mm}rccrr}
\hline
Run & UT date & Start time & Exp. & Points & Length\\
no. & (2007) & (BJD-2\,450\,000) & (s) &  & (h)\\
\hline
01 & Jan 15 & 4115.614 & 30 & 299 & 2.81\\
02 & Jan 17 & 4117.622 & 30 & 154 & 1.54\\
03 & Jan 26 & 4126.615 & 30 & 233 & 2.19\\
04 & Jan 28 & 4128.544 & 30 & 441 & 4.17\\
05 & Jan 30 & 4130.528 & 30 & 493 & 4.58\\
06 & Feb 17 & 4148.551 & 30 & 375 & 3.53\\
07 & Mar 15 & 4175.283 & 30 & 725 & 9.27\\
08 & Mar 16 & 4176.279 & 30 & 960 & 9.29\\
09 & Mar 22 & 4182.356 & 30 & 562 & 6.04\\
10 & Mar 24 & 4184.496 & 30 & 410 & 3.76\\
11 & Mar 25 & 4185.398 & 30 & 338 & 3.12\\
12 & Mar 26 & 4186.282 & 30 & 636 & 5.88\\
13 & Mar 27 & 4187.273 & 30 & 718 & 7.94\\
14 & Mar 30 & 4190.307 & 30 & 862 & 8.05\\
15 & Mar 31 & 4191.357 & 30 & 325 & 2.98\\
16 & Apr 01 & 4192.289 & 30 & 596 & 5.47\\
17 & Apr 03 & 4194.276 & 30 & 202 & 1.92\\
18 & Apr 12 & 4203.388 & 10 & 1483 & 5.58\\
19 & Apr 13 & 4204.337 & 10 & 1808 & 6.80\\
20 & Apr 14 & 4205.304 & 10 & 1946 & 7.66\\
21 & Apr 15 & 4206.309 & 10 & 1608 & 7.27\\
22 & Apr 16 & 4207.296 & 10 & 1422 & 5.51\\
23 & Apr 17 & 4208.378 & 10 & 1439 & 5.26\\
24 & May 10 & 4231.316 & 10 & 1026 & 4.08\\
25 & May 12 & 4233.358 & 30 & 259 & 2.33\\
26 & May 13 & 4234.315 & 10 & 1192 & 4.32\\
27 & May 14 & 4235.372 & 10 & 763 & 2.82\\
28 & May 16 & 4237.365 & 30 & 342 & 3.12\\
\multicolumn{2}{l}{Total:} & & \multicolumn{2}{r}{21\,617} & 137.27\\
\hline
\end{tabular}
\end{table}

\begin{figure*}
\centering
\includegraphics[width=16cm]{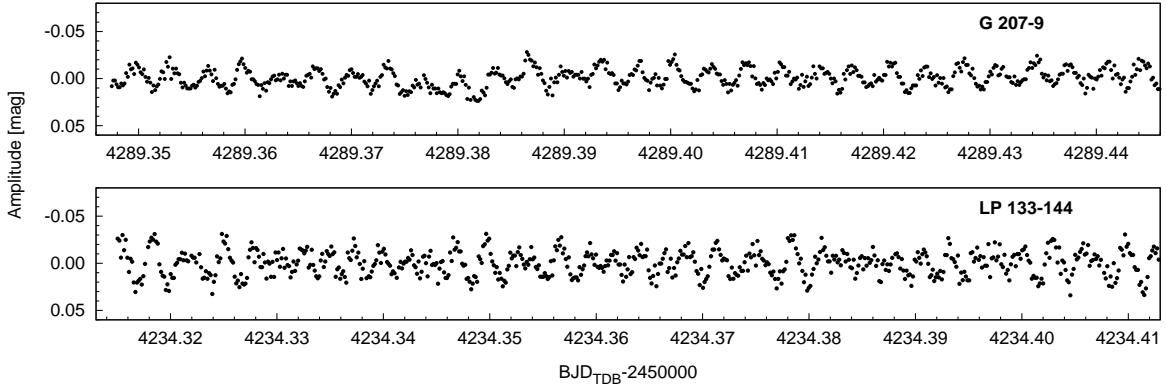}
\caption{Segments of the light curves of G~207-9 (\textit{upper panel}) and LP~133-144 (\textit{lower panel}) 
obtained at Piszk\'estet\H o using 10\,s integration times.}
\label{fig:lcshort}
\end{figure*}

\begin{figure*}
\includegraphics[width=6cm]{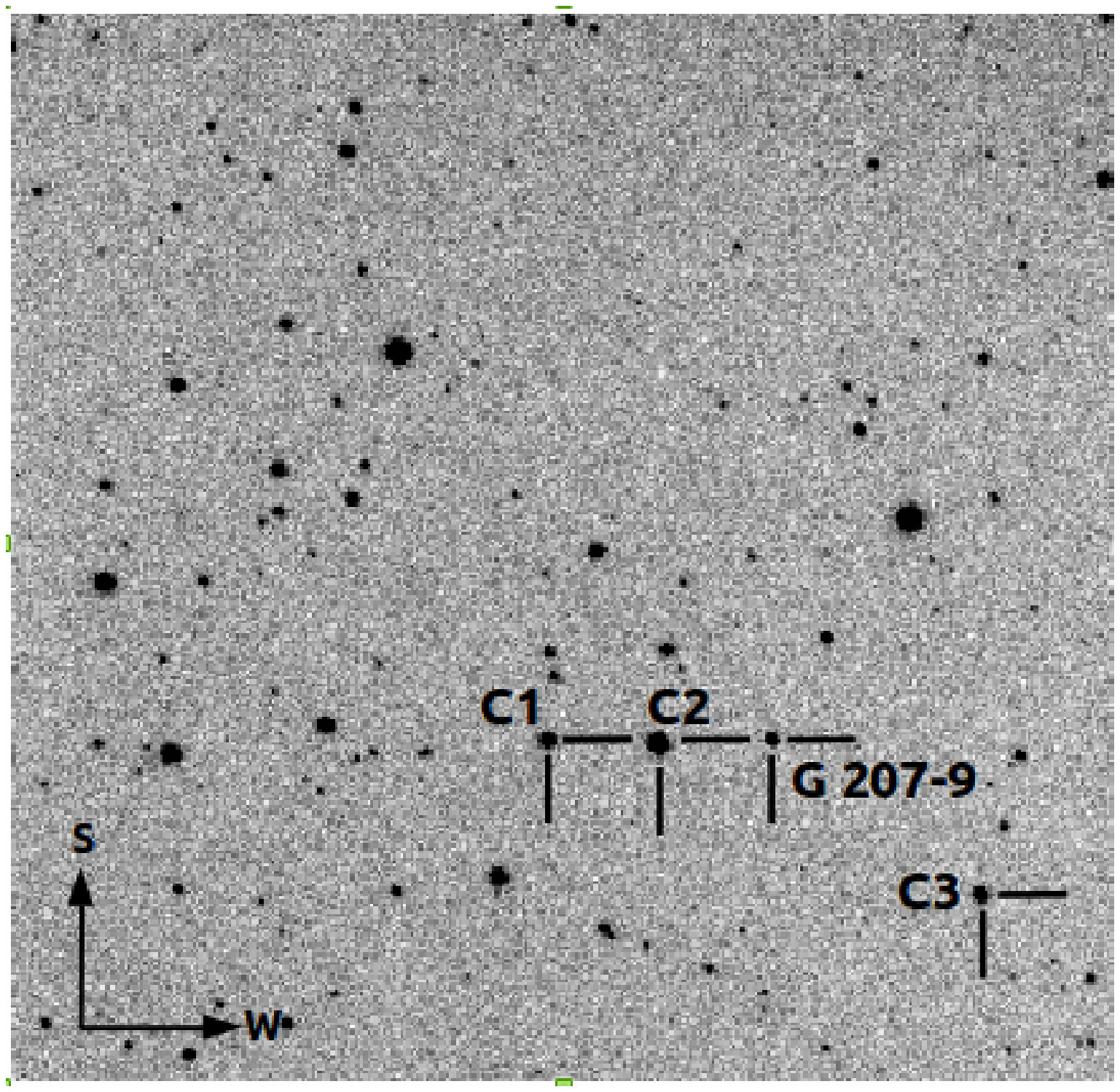}
\includegraphics[width=6cm]{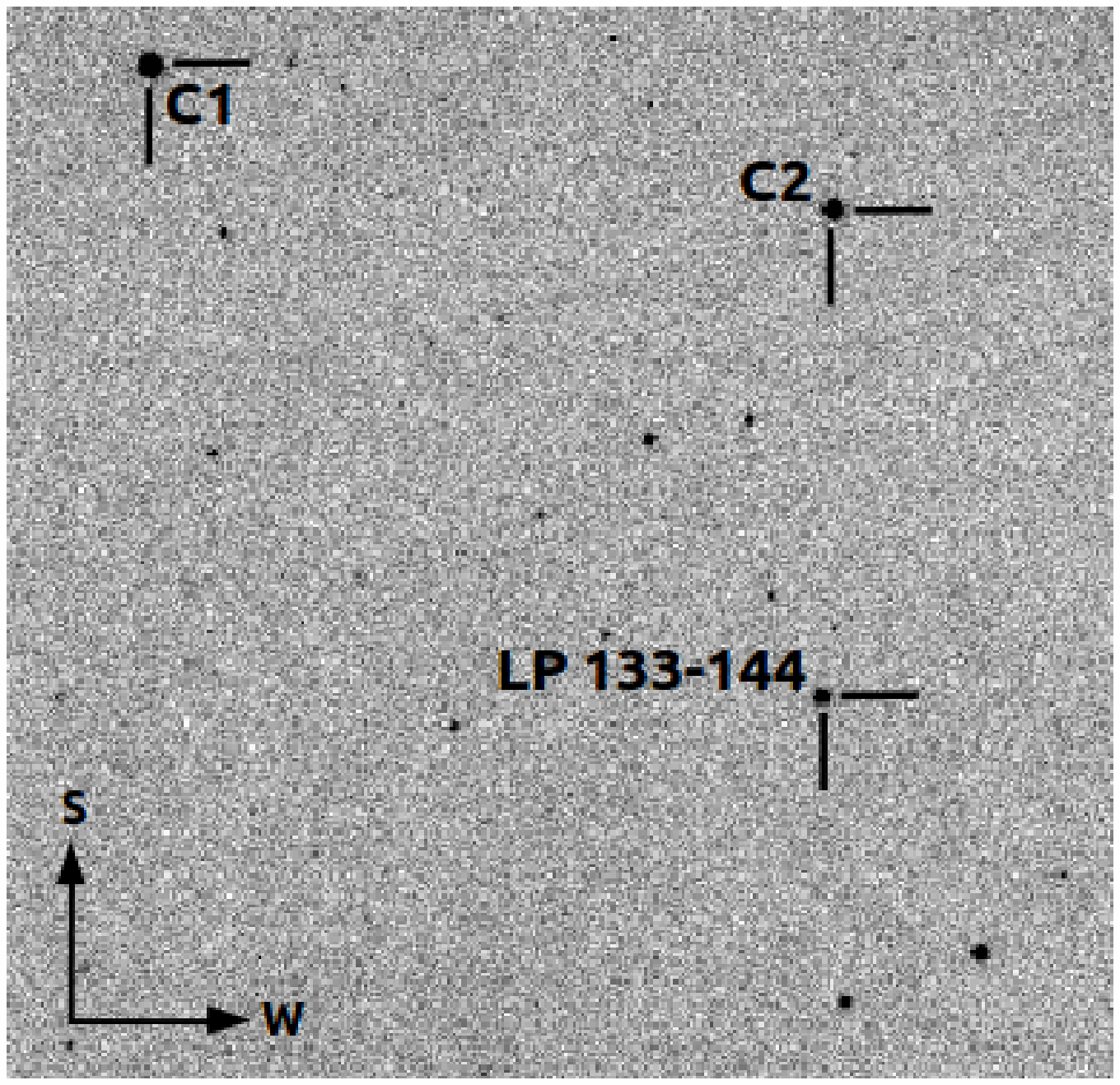}
\caption{CCD frames with the variable and comparison stars marked. The field of view is $\approx7\arcmin\times7\arcmin$.}
\label{fig:ccd}
\end{figure*}

We reduced the raw data frames following the standard procedure: we applied bias, dark and flat corrections on the frames 
using \textsc{iraf}\footnote{\textsc{iraf} is distributed by the National Optical Astronomy Observatories, which are 
operated by the Association of Universities for Research in Astronomy, Inc., under cooperative agreement with the 
National Science Foundation.} routines, and performed aperture photometry of the variable and comparison stars with 
the \textsc{iraf daophot} package. We converted the observational times of every data point to barycentric Julian dates
in barycentric dynamical time ($\mathrm{BJD_{TDB}}$) using the applet of 
\citet{2010PASP..122..935E}\footnote{http://astroutils.astronomy.ohio-state.edu/time/utc2bjd.html}.
We then checked the comparison star candidates for variability and instrumental effects. We selected three stars 
in the field of G~207-9 and two stars in the field of LP~133-144 and used the averages of these reference stars as
comparisons for the differential photometry of the two pulsators. The panels of Fig.~\ref{fig:ccd} show the 
variable and the comparison stars in the CCD fields. We applied low-order polynomial fits
to the light curves to correct for the instrumental trends and for the atmospheric extinction. This method did not 
affect the pulsation frequency domains. 
Figure~\ref{fig:lcshort} shows two illustrative light curve segments of G~207-9 and LP~133-144.
All the light curves obtained for both pulsators are presented in Appendix~\ref{app:g207} and in Appendix~\ref{app:lp133}.

\section{Frequency analyses of the light curves}

We determined the frequency content of the datasets on daily, weekly or monthly, and yearly time bases.
We analysed the daily observations with custom developed software tools, as the command-line light curve fitting 
program \textsc{LCfit} \citep{2012KOTN...15....1S}. \textsc{LCfit} has linear (amplitudes and phases) and nonlinear 
(amplitudes, phases and frequencies) least-squares fitting options, utilizing an implementation of the
Levenberg-Marquardt least-squares fitting algorithm. The program can handle unequally spaced and gapped datasets.
\textsc{LCfit} is scriptable easily, which made the analysis of the relatively large number of nightly datasets
very effective.

We performed the standard Fourier analyses of the weekly or monthly data subsets and the whole light curves with the 
photometry modules of the Frequency Analysis and Mode Identification for Asteroseismology (\textsc{famias}) software 
package \citep{2008CoAst.155...17Z}. Following the traditional way, we accepted a frequency peak as significant if
its amplitude reached the 4 signal-to-noise ratio (S/N). The noise level was calculated as the average amplitude
in a $\pm1200\,\mu$Hz interval around the given frequency.

\subsection{G~207-9}

\subsubsection{Previous observations}

G~207-9 was announced as the 8th known member of pulsating white dwarf stars in 1976 \citep{1976ApJ...207L..37R}.
Four high ($\mathrm{F}_1$--$\mathrm{F}_4$) and one low ($\mathrm{F}_5$) amplitude peaks were detected
at $\mathrm{F}_1=1354$, $\mathrm{F}_2=1794$, $\mathrm{F}_3=3145$, $\mathrm{F}_4=3425$ 
and $\mathrm{F}_5=3860\,\mu$Hz.
Even though G~207-9 is a relatively bright target, and has been known as a pulsator for decades, no other time series 
photometric observations and frequency analysis have been published on this star up to now.

\subsubsection{Konkoly observations}
\label{sect:g207freq}

\begin{figure}
\centering
\includegraphics[width=\columnwidth]{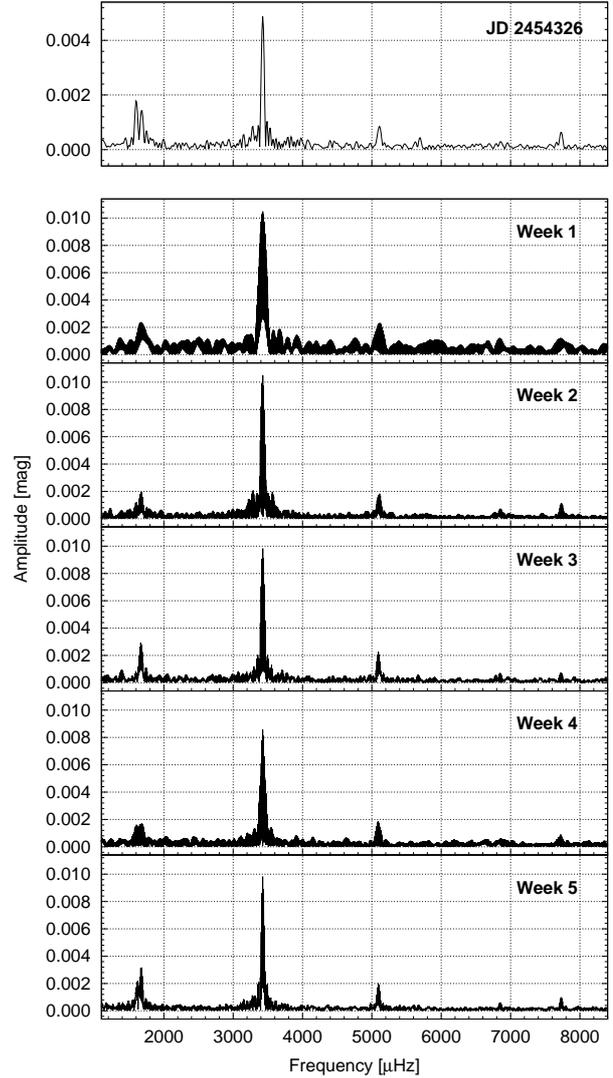}
\caption{G~207-9: amplitude spectra of one night's observation (\textit{top panel}) and the weekly datasets (\textit{lower panels}).}
\label{fig:g207FTa}
\end{figure}

The analyses of the daily datasets revealed one dominant and four low amplitude peaks in the FTs.
The dominant frequency of all nights' observations in 2007 was at $3426\,\mu$Hz. This was the 3rd
highest amplitude mode in 1975 ($\mathrm{F}_4=3425\,\mu$Hz).
Two lower amplitude frequencies
at $1672$ and $5098\,\mu$Hz reached the 4\,S/N detection limit in 12 and 14 of the daily datasets, respectively.
Two additional low-amplitude frequencies were detected at $1608$ and $7725\,\mu$Hz in 3 and 4 cases, respectively.
These frequencies are medians of the daily values.
The $1608$ and $1672\,\mu$Hz frequencies could be determined separately only in the last three nights' datasets.
We present the FT of one night's dataset (the second longest run) in the first panel of Fig.~\ref{fig:g207FTa}.

Considering the consecutive nights of observations, five weekly time base datasets can be formed: 
Week~1 (JD\,2\,454\,185--193), Week~2 (JD\,2\,454\,267--272), Week~3 (JD\,2\,454\,288--292),
Week~4 (JD\,2\,454\,308--314) and Week~5 (JD\,2\,454\,323--328). The Fourier analyses of these
data verified the five frequencies found by the daily observations. In three cases the first harmonic
of the dominant frequency was also detected. Additionally, the analysis of the Week~2 and Week~5 data
suggested that the peaks at $\sim1672$ and $\sim5098\,\mu$Hz may be actually doublets or triplets and not
singlet frequencies. The separations of the frequency components were found to be between $\sim2$ (close to 
the resolution limit) and $14\,\mu$Hz, but we mark these findings uncertain because of the effect of the 1\,d$^{-1}$ aliasing.
The 2nd--6th panels of Fig.~\ref{fig:g207FTa} shows the FTs of the weekly datasets. We found only slight amplitude
variations from one week to another. The amplitude of the dominant frequency varied between 8.6 and 10.5\,mmag.

The standard pre-whitening of the whole dataset resulted 26 frequencies above the 4\,S/N limit. Most of them are 
clustering around the frequencies already known by the analyses of the daily and weekly datasets.
Generally, amplitude and (or) phase variations during the observations can be responsible for
the emergence of such closely spaced peaks. In such cases, these features are just artefacts in the FT, as we fit the light
curve with fixed amplitudes and frequencies during the standard pre-whitening process.
Another possibility is that some of the closely spaced peaks are rotationally split frequencies. We can 
resolve such frequencies if the time base of the observations is long enough. The Rayleigh frequency resolution 
($1/\Delta T$) of the whole dataset is $0.08\,\mu$Hz.
We also have to consider the 1\,d$^{-1}$ alias problem of single-site observations, which results uncertainties in the
frequency determination.

\begin{table}
\centering
\caption{G~207-9: frequency content of the 2007 dataset. The errors were calculated by Monte Carlo simulations.
$\delta f$ denotes the frequency differences of the closely spaced frequencies to $f_1$, $f_2$ or $f_4$.}
\label{table:g207freq}
\begin{tabular}{p{3.2mm}ccrrr}
\hline
 & \multicolumn{1}{c}{Frequency} & \multicolumn{1}{c}{Period} & \multicolumn{1}{c}{$\lvert\delta f\rvert$} & \multicolumn{1}{c}{Ampl.} & \multicolumn{1}{c}{S/N}\\
 & & & & \multicolumn{1}{c}{$\pm0.1$} & \\
 & \multicolumn{1}{c}{($\mu$Hz)} & \multicolumn{1}{c}{(s)} & \multicolumn{1}{c}{($\mu$Hz)} & \multicolumn{1}{c}{(mmag)} & \\
\hline
$f_1$ & $3426.303\pm0.001$ & 291.9 & & 10.1 & 111.5\\
$f_2$ & $1678.633\pm0.003$ & 595.7 & & 2.0 & 15.4\\
$f_1^-?$ & $3414.639\pm0.004$ & 292.9 & 11.7 & 1.6 & 18.0\\
$f_3$ & $5098.861\pm0.003$ & 196.1 & & 1.2 & 12.5\\
$f_2^-?$ & $1667.328\pm0.005$ & 599.8 & 11.3 & 1.1 & 8.6\\
$f_4$ & $1603.071\pm0.004$ & 623.8 & & 1.1 & 8.1\\
$f_1^+?$ & $3437.384\pm0.005$ & 290.9 & 11.1 & 1.0 & 11.6\\
$f_5$ & $7726.540\pm0.003$ & 129.4 & & 1.0 & 13.0\\
$f_4^-?$ & $1595.481\pm0.004$ & 626.8 & 7.6 & 0.8 & 6.1\\
$2f_1$ & $6852.604\pm0.005$ & 145.9 & & 0.6 & 8.5\\
$f_6$ & $3146.670\pm0.007$ & 317.8 & & 0.5 & 5.0\\
$f_7$ & $3276.485\pm0.008$ & 305.2 & & 0.4 & 4.6\\
\hline
\end{tabular}
\end{table}

\begin{figure*}
\centering
\includegraphics[width=17.5cm]{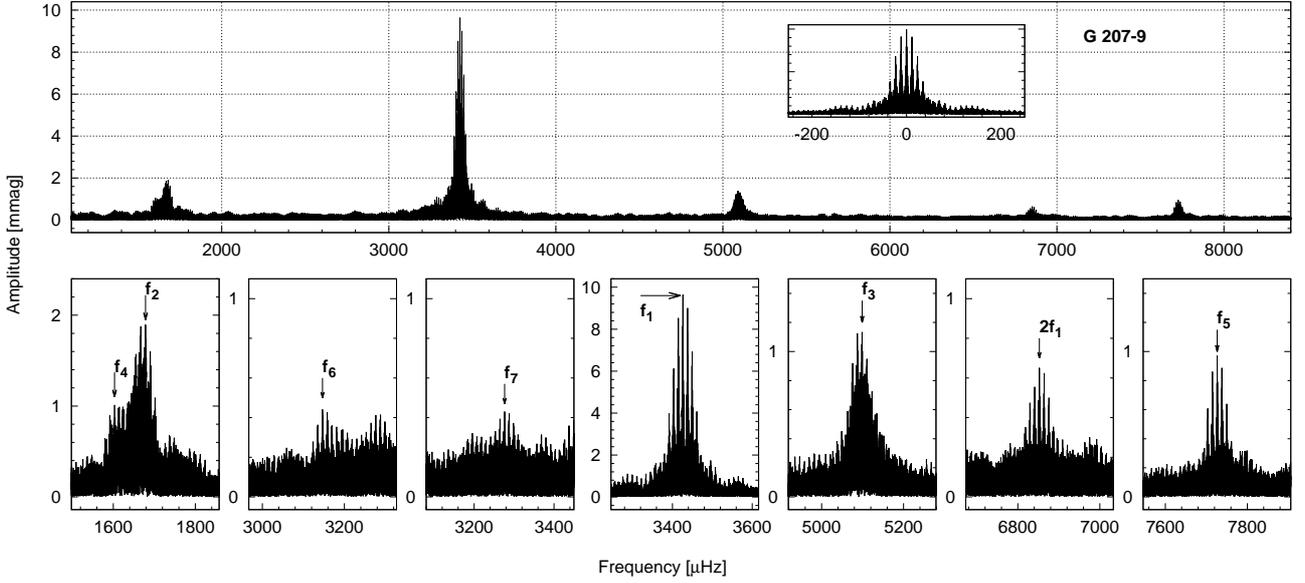}
\caption{G~207-9: FT of the whole dataset, the window function is given in the inset (\textit{top panel}).
We emphasized the frequency domains of $f_1-f_7$ and $2f_1$, as seen in the original FT or during the pre-whitening 
process (\textit{lower panels}).}
\label{fig:g207prewh}
\end{figure*}

\begin{figure}
\centering
\includegraphics[width=\columnwidth]{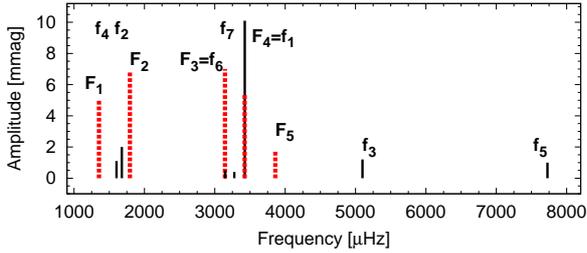}
\caption{G~207-9: comparison of the frequencies obtained in 1975 (\textit{red dashed lines}) and in 2007 (\textit{black solid lines}).
The amplitudes of the 1975 observations are from the paper of \citet{2006ApJ...640..956M}.}
\label{fig:g207oldnew}
\end{figure}

We checked the frequency content of the whole dataset by averaging three consecutive data points of the $10\,$s measurements
as a test. That is, we created a new, more homogeneous dataset mimicking $30\,$s exposure times. We then compared the frequency solutions
of this $30\,$s dataset with the frequencies of the original mixed $10$--$30\,$s data. Finally, we accepted as the frequencies
characterizing the whole light curve the frequencies that could be determined in both datasets, that is, without 1\,d$^{-1}$ differences.
This resulted a reduced frequency list of 12 frequencies. We list them in Table~\ref{table:g207freq}. There are still several 
closely spaced frequencies around three of the main frequencies ($f_1$, $f_2$ and $f_4$) remained with separations between $7.6$ and $11.7\,\mu$Hz. 
In the case of $f_1$ and $f_2$, these separations are close to $11.574\,\mu$Hz (1\,d$^{-1}$). 
It is possible that at least some of these frequency components are results of rotational splitting, but considering
the uncertainties mentioned above, we do not accept them as rotationally split frequencies.
In such cases when the frequency separations of the rotationally split components are around 1\,d$^{-1}$, multi-site or space-based 
observations are needed for reliable determination of the star's rotational rate.

Summing it up: besides the five frequencies ($f_1$--$f_5$) known also by the analyses of shorter (daily and weekly) data segments, 
we could detect two additional independent frequencies ($f_6$ and $f_7$) in the whole dataset. Frequency $f_6$ 
at $3146.7\,\mu$Hz was also detected in 1975 ($\mathrm{F}_3=3145\,\mu$Hz). Moreover,
this was one of the dominant peaks at that time. Frequency $f_7$ at $3276.5\,\mu$Hz is a newly detected one.
Note that the frequency $f_7$ is close to $f_2+f_4=3281.7\,\mu$Hz,
however, the difference is $5.2\,\mu$Hz, which seems too large to claim that
$f_7$ is the linear combination of these peaks considering the errors. Thus, we consider $f_7$ as an independent mode.
Fig.~\ref{fig:g207prewh} shows the FT of the whole dataset and the frequency domains of $f_1-f_7$
on separate panels. 

Comparing the frequency content of the 1975 and 2007 observations, we can 
conclude that three of the five frequencies found in the 1975 dataset did not appear in 2007 ($\mathrm{F}_1$, 
$\mathrm{F}_2$ and $\mathrm{F}_5$), while two stayed at an observable level 
($\mathrm{F}_3=f_6$ and $\mathrm{F}_4=f_1$). Figure~\ref{fig:g207oldnew} summarizes the frequencies of the
two epochs. It seems that even though there were no large amplitude variations during our five-months observing season in 2007,
on the time scale of years or decades, remarkable changes can happen in the pulsation of G~207-9: new frequencies can
be excited to a significant level, while other modes can disappear.

\subsection{LP~133-144}

\subsubsection{Previous observations}

The variability of LP~133-144 was discovered in 2003 \citep{2004ApJ...600..404B}.
Four pulsation frequencies were determined at that time, including two closely spaced peaks: $\mathrm{F}_1=3055.1$, $\mathrm{F}_2=3258.4$, 
$\mathrm{F}_3=3284.1$ and $\mathrm{F}_4=4780.6\,\mu$Hz.
Similarly to the case of G~207-9, no further results of time series photometric observations have been published up to now.  

\subsubsection{Konkoly observations}
\label{sect:lp133freq}

We found four recurring frequencies in the daily datasets at 3055, 3270, 3695 and 4780\,$\mu$Hz (median values). Their amplitudes varied 
from night to night, but the 4780\,$\mu$Hz peak was the dominant in almost all cases. One additional peak exceeded the 4\,S/N limit
at 5573\,$\mu$Hz, but on one night only.

We created four monthly datasets and analysed them independently. These are Month~1 (JD\,2\,454\,115--130), Month~2 (JD\,2\,454\,175--194), 
Month~3 (JD\,2\,454\,203--208) and Month~4 (JD\,2\,454\,231--237). The analyses of the monthly data revealed that at the 3270, 
3695 and 4780\,$\mu$Hz frequencies there are actually doublets or triplets with 2.6--4.7\,$\mu$Hz frequency separations. 
This explains the different amplitudes in the daily FTs.
The 3055\,$\mu$Hz frequency was found to be a singlet. In Month~3, the linear combination of the largest amplitude components of the 3270 and 
4780\,$\mu$Hz multiplets also could be detected. The 5573\,$\mu$Hz frequency was significant in Month~2.

\begin{figure}
\centering
\includegraphics[width=\columnwidth]{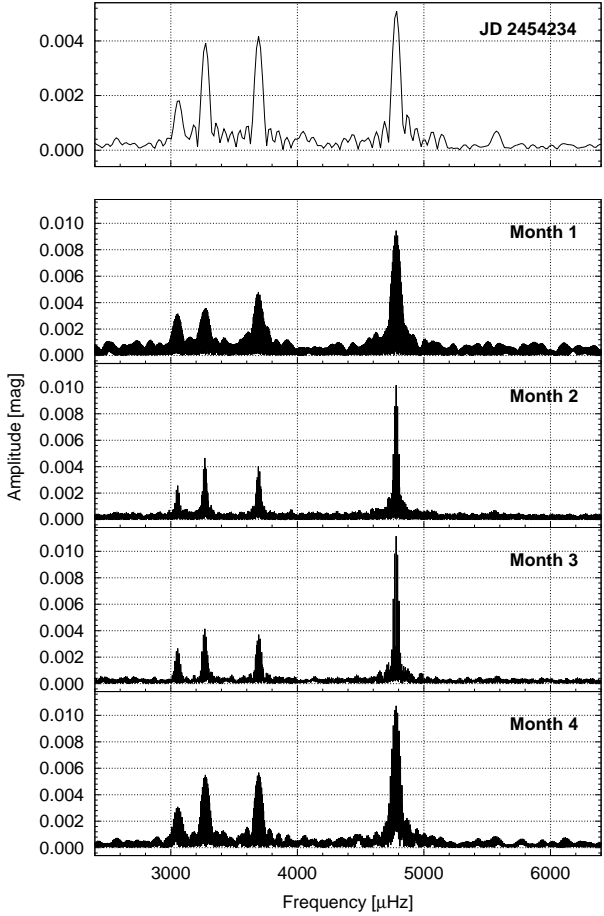}
\caption{LP~133-144: amplitude spectra of one night's observation (\textit{top panel}) and the monthly datasets (\textit{lower panels}).}
\label{fig:lp133FTa}
\end{figure}

The panels of Fig.~\ref{fig:lp133FTa} show the FT of one daily dataset and the monthly data. As in the case of G~207-9, there were no
remarkable amplitude variations from one month to another.

The analysis of the whole 2007 dataset resulted in the detection of 19 significant frequencies in the $\sim2300-8000\,\mu$Hz frequency
region. We also performed the test analysis utilizing the averaged 30\,s dataset, which confirmed the presence of the 14 largest amplitude
frequencies (the other five peaks remained slightly under the significance level). Thus we accepted them as the frequencies 
characterizing the pulsation of LP~133-144 and list them in Table~\ref{table:lp133freq}.
The Rayleigh frequency resolution of the whole dataset is $0.09\,\mu$Hz.

\begin{table}
\centering
\caption{LP~133-144: frequency content of the 2007 dataset. The errors were calculated by Monte Carlo simulations.
$\delta f$ denotes the frequency differences of the closely spaced frequencies to $f_1$, $f_2$ $f_3$ or $f_6$.
We discuss the case of $f_6$ in the text.
The signal-to-noise ratios refer to the original 10-30\,s dataset.}
\label{table:lp133freq}
\begin{tabular}{p{3.2mm}ccrrr}
\hline
 & \multicolumn{1}{c}{Frequency} & \multicolumn{1}{c}{Period} & \multicolumn{1}{c}{$\lvert\delta f\rvert$} & \multicolumn{1}{c}{Ampl.} & \multicolumn{1}{c}{S/N}\\
 & & & & \multicolumn{1}{c}{$\pm0.1$} & \\
 & \multicolumn{1}{c}{($\mu$Hz)} & \multicolumn{1}{c}{(s)} & \multicolumn{1}{c}{($\mu$Hz)} & \multicolumn{1}{c}{(mmag)} & \\
\hline
$f_1$ & 4780.555$\pm$0.001 & 209.2 & & 10.9 & 100.9\\ 
$f_2$ & 3269.302$\pm$0.001 & 305.9 & & 3.9 & 35.4\\
$f_3$ & 3695.083$\pm$0.002 & 270.6 & & 3.5 & 31.2\\
$f_3^-$ & 3691.627$\pm$0.002 & 270.9 & 3.5 & 3.4 & 30.5\\
$f_2^+$ & 3272.475$\pm$0.002 & 305.6 & 3.2 & 3.0 & 26.7\\
$f_4$ & 3055.125$\pm$0.002 & 327.3 & & 2.8 & 25.1\\
$f_3^+$ & 3698.551$\pm$0.003 & 270.4 & 3.5 & 2.0 & 18.4\\
$f_2^-$ & 3266.125$\pm$0.005 & 306.2 & 3.2 & 1.2 & 10.4\\
$f_1^+$ & 4784.696$\pm$0.005 & 209.0 & 4.1 & 1.1 & 10.6\\
$f_1^-$ & 4776.400$\pm$0.007 & 209.4 & 4.2 & 1.0 & 8.7\\
$f_5$ & 7116.986$\pm$0.010 & 140.5 & & 0.6 & 5.9\\
$f_6^+$ & 5574.381$\pm$0.009 & 179.4 & 4.8 & 0.6 & 6.0\\
$2f_1$ & 9561.115$\pm$0.011 & 104.6 & & 0.5 & 5.5\\
$f_6^-$ & 5564.876$\pm$0.013 & 179.7 & 4.7 & 0.5 & 5.1\\
($f_6$) & 5569.618$\pm$0.020 & 179.5 & & 0.4 & 4.1\\
\hline
\end{tabular}
\end{table}

\begin{figure*}
\centering
\includegraphics[width=17.5cm]{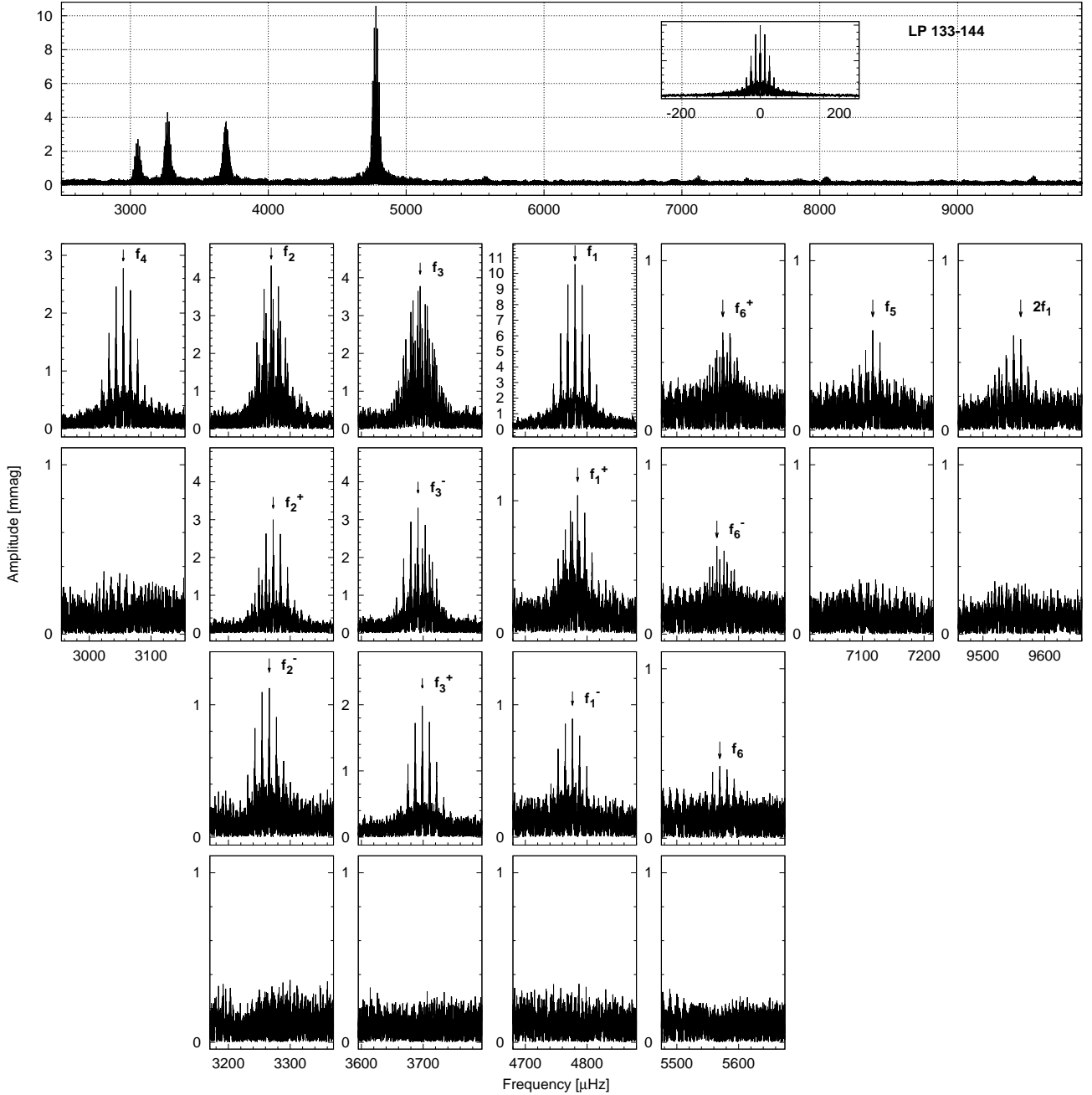}
\caption{LP~133-144: FT of the whole dataset, the window function is given in the inset (\textit{top panel}).
We also plotted the FTs of the consecutive pre-whitening steps at the multiplet frequencies $f_1$, $f_2$, $f_3$ and $f_6$,
the frequency domains of the singlets $f_4$ and $f_5$, and the peak at $2f_1$ (\textit{lower panels}). The \textit{bottom panels}
show the residual spectra after pre-whitening with the denoted frequencies.}
\label{fig:lp133prewh}
\end{figure*}

The first eleven peaks in Table~\ref{table:lp133freq} are three triplets with frequency separations of $4.1-4.2\,\mu$Hz ($f_1$),  $3.2\,\mu$Hz ($f_2$)
or $3.5\,\mu$Hz ($f_3$), and two singlet frequencies ($f_4$ and $f_5$). In the case of $f_6$, three
peaks can be determined in the original 10-30\,s dataset with frequency separations of $4.7-4.8\,\mu$Hz. However, the low amplitude central 
peak of this triplet at $f_6=5569.6\,\mu$Hz do not reach the 4\,S/N significance limit in the test 30\,s data.
Still, to make the discussion of the triplet structures clear, we added $f_6$ to the list of Table~\ref{table:lp133freq} in parentheses.
Besides these, the first harmonic of $f_1$ also appeared. Fig.~\ref{fig:lp133prewh} shows the FT of the whole dataset, 
the consecutive pre-whitening steps at the multiplet frequencies and at the 
frequency domains of $f_4$, $f_5$ and $2f_1$. 

\begin{figure}
\centering
\includegraphics[width=\columnwidth]{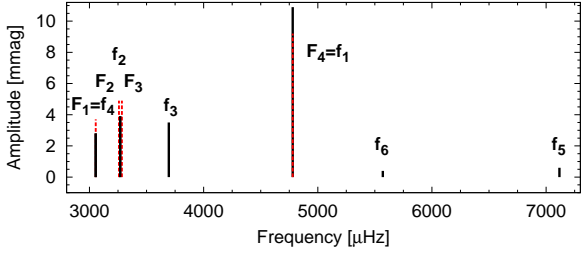}
\caption{LP~133-144: comparison of the frequencies obtained in 2003 (\textit{red dashed lines}) and in 2007 (\textit{black solid lines}).}
\label{fig:lp133oldnew}
\end{figure}

We plot the frequencies of \citet{2004ApJ...600..404B} and the frequencies found in the 2007 Konkoly observations together in 
Fig.~\ref{fig:lp133oldnew}. Assuming that the closely spaced peaks at $\mathrm{F}_2$ and $\mathrm{F}_3$ are results of the not
properly resolved components of the $f_2$ triplet, we found, with similar amplitudes, all the frequencies observed in 2003. 
Besides these, we detected three new frequencies: a relatively large amplitude mode at $f_3$, and two additional low-amplitude 
modes at $f_5$ and $f_6$. That is, we doubled the number of modes can be used for the asteroseismic fits.

\begin{figure}
\centering
\includegraphics[width=\columnwidth]{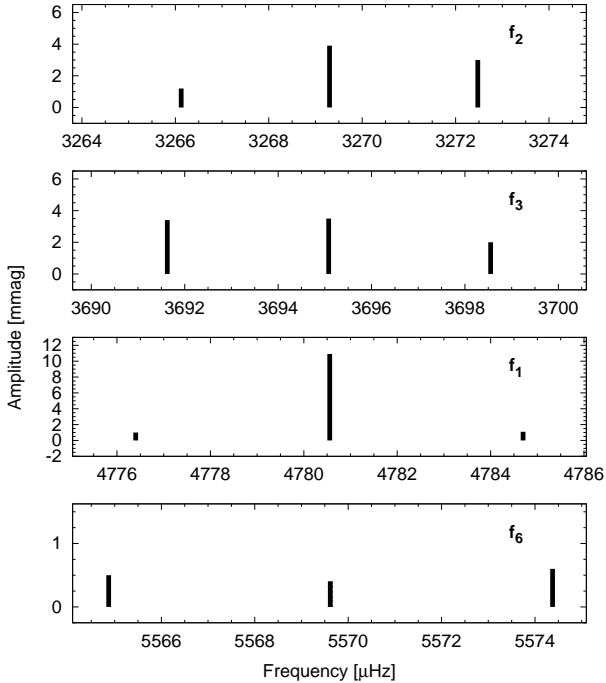}
\caption{LP~133-144: schematic plot of the triplets found at different frequency domains.
The frequency errors are comparable to the width of the lines.}
\label{fig:lp133triplet}
\end{figure}

The schematic plot of the triplets can be seen in Fig.~\ref{fig:lp133triplet}.
It is clearly visible that the frequency separations of the components are larger at higher frequencies.
We discuss the rotation of LP~133-144 based on the investigation of these triplets in Sect.~\ref{sect:lp133rot}.

\section{Asteroseismology}

We built our model grid for the asteroseismic investigations of our targets utilizing the White Dwarf Evolution Code (\textsc{wdec}; 
\citealt{1974PhDT........56L, 1969ApJ...156.1021K, 1975ApJ...200..306L, 
1991PhDT.........XX, 1986PhDT.........2K, 1990PhDT.........5W, 1993PhDT.........4B, 1998PhDT........21M, 2008ApJ...675.1512B}).
The \textsc{wdec} evolves a hot polytrope model ($\sim10^5$\,K) down to the requested temperature, and provides an equilibrium,
thermally relaxed solution to the stellar structure equations. Then we are able to calculate the set of possible zonal ($m=0$) pulsation modes
according to the adiabatic equations of non-radial stellar oscillations \citep{1989nos..book.....U}.
We utilized the integrated evolution/pulsation form of the \textsc{wdec} code created by \citet{2001PhDT.........1M} to derive the
pulsation periods for the models with the given stellar parameters. More details on the physics applied in the \textsc{wdec}
can be found with references in \citet{2008ApJ...675.1512B} and in our previous papers on two ZZ Ceti stars \citep{2009MNRAS.399.1954B,
2013MNRAS.432..598P}.

Considering the limited visibility of high spherical degree ($l$) modes due to geometric cancellation effects, we calculated
the periods of dipole ($l=1$) and quadrupole ($l=2$) modes for the model stars only. The goodness of the fit between the observed ($P_i^{\rmn{obs}}$)
and calculated ($P_i^{\rmn{calc}}$) periods was characterized by the root mean square ($\sigma_\mathrm{{rms}}$) value calculated for every 
model with the \textsc{fitper} program of \citet{2007PhDT........13K}:

\begin{equation}
\sigma_\mathrm{{rms}} = \sqrt{\frac{\sum_{i=1}^{N} (P_i^{\rmn{calc}} - P_i^{\rmn{obs}})^2}{N}}
\label{equ1}
\end{equation}

\noindent where \textit{N} is the number of observed periods. 

We varied five main stellar parameters to build our model grid: 
the effective temperature ($T_{\rmn{eff}}$), the stellar mass ($M_*$),
the mass of the hydrogen layer ($M_\rmn{H}$), the central oxygen abundance ($X_\rmn{O}$) and the fractional mass point where the
oxygen abundance starts dropping ($X_{\rmn{fm}}$). We fixed the mass of the helium layer ($M_\rmn{He}$) at $10^{-2}\,M_*$.
The grid covers the parameter range $11\,400-12\,800$\,K in $T_{\rmn{eff}}$ (the middle and hot part of the ZZ Ceti instability strip), 
$0.500-0.900\,M_{\sun}$ in stellar mass, $10^{-4}-10^{-8}\,M_*$ in $M_\rmn{H}$, $0.3-0.9$ in $X_\rmn{O}$ and $0.1-0.7$ in $X_{\rmn{fm}}$. 
We used step sizes of $200$\,K ($T_{\rmn{eff}}$), $0.005\,M_{\sun}$ ($M_*$), $0.2$\,dex (log\,$M_\rmn{H}$) and 
0.1 ($X_\rmn{O}$ and $X_{\rmn{fm}}$).

\subsection{Period lists}

In the case of G~207-9, we could detect seven linearly independent pulsation frequencies by the 2007 Konkoly dataset ($f_1-f_7$; 
see Table~\ref{table:g207freq}).
The question is, if we could add more frequencies to this list by the 1975 observations of \citet{1976ApJ...207L..37R}.
As we mentioned already in Sect.~\ref{sect:g207freq}, two of the frequencies detected in 1975 were also found in the
Konkoly data ($\mathrm{F}_3=f_6$ and $\mathrm{F}_4=f_1$). The status of the remaining three 1975 frequencies is
questionable. Assuming at least a couple of $\mu$Hz errors for the 1975 frequencies, $\mathrm{F}_1=\mathrm{F}_3-\mathrm{F}_2$
(or $\mathrm{F}_2=\mathrm{F}_3-\mathrm{F}_1$, or $\mathrm{F}_3=\mathrm{F}_1+\mathrm{F}_2$), thus, these three frequencies
do not seem to be linearly independent. The fact that $\mathrm{F}_2$ and $\mathrm{F}_3$ are the two dominant peaks in the FT of 
\citet{1976ApJ...207L..37R} suggests that $\mathrm{F}_2$ and $\mathrm{F}_3$ might be the parent modes and $\mathrm{F}_1$
is a combination peak. Furthermore, 
\citet{1976ApJ...207L..37R} pointed out that $\mathrm{F}_5-\mathrm{F}_4\approx\mathrm{F}_2-\mathrm{F}_1$, thus, further
combinations are possible. We also note that $f_5$ of the Konkoly dataset is almost at twice the value of $\mathrm{F}_5$
($\delta f=6.5\mu$Hz), however, there is no sign of any pulsation frequency at $0.5f_5$ in the 2007 data.

We used two sets of observed periods to fit the calculated ones. One set consists of the seven periods of $f_1-f_7$
observed in 2007, while we complemented this list with the period of $\mathrm{F}_2$ detected in 1975 to create another set. 
We selected $\mathrm{F}_2$
because it was the second largest amplitude peak in 1975, which makes it a good candidate for an additional normal mode.

In LP~133-144, we found all the previously observed frequencies in our 2007 dataset, as we show in Sect.~\ref{sect:lp133freq}.
Thus, we cannot add more frequencies to our findings, and performed the model fits with six periods. We summarized the periods
utilized for modelling in Table~\ref{table:periods} for both stars.

\begin{table}
\centering
\caption{G~207-9 and LP~133-144: periods utilized for the model fits.}
\label{table:periods}
\begin{tabular}{p{3.2mm}cp{3.2mm}c}
\hline
\multicolumn{2}{l}{G~207-9} & \multicolumn{2}{l}{LP~133-144}\\
 & Period & & Period\\
 & (s) & & (s)\\
\hline
$f_1$ & 291.9 & $f_1$ & 209.2\\ 
$f_2$ & 595.7 & $f_2$ & 305.9\\
$f_3$ & 196.1 & $f_3$ & 270.6\\
$f_4$ & 623.8 & $f_4$ & 327.3\\
$f_5$ & 129.4 & $f_5$ & 140.5\\
$f_6$ & 317.8 & $f_6$ & 179.5\\
$f_7$ & 305.2 & & \\
+$\mathrm{F}_2$ & 557.4 & & \\ 
\hline
\end{tabular}
\end{table}

\subsection{Best-fitting models for G~207-9}

We determined the best-matching models considering several cases: at first, we let all modes to be either $l=1$ or $l=2$. 
Then we assumed that the dominant peak is an $l=1$, considering the better visibility of $l=1$ modes over $l=2$ ones.
At last, we searched for the best-fitting models assuming that at least four of the modes is $l=1$, including the 
dominant frequency.

We obtained the same model as the best-fitting asteroseismic solution both for the seven- and eight-period fits.
It has $T_{\rmn{eff}}=12\,000$\,K, $M_*=0.870\,M_{\sun}$ and $M_\rmn{H}=10^{-4}\,M_*$. This model has the lowest
$\sigma_\mathrm{{rms}}$ ($1.04-1.06$\,s) both if we do not apply any restrictions on the $l$ values of the modes, and as it gives
$l=1$ solution to the dominant frequency, this model is also the best-fit if we assume that the $291.9$\,s mode
is $l=1$. Note that in this model solution only this mode is an $l=1$, all the other six or seven modes are $l=2$. 

\begin{figure*}
\centering
\includegraphics[width=17.5cm]{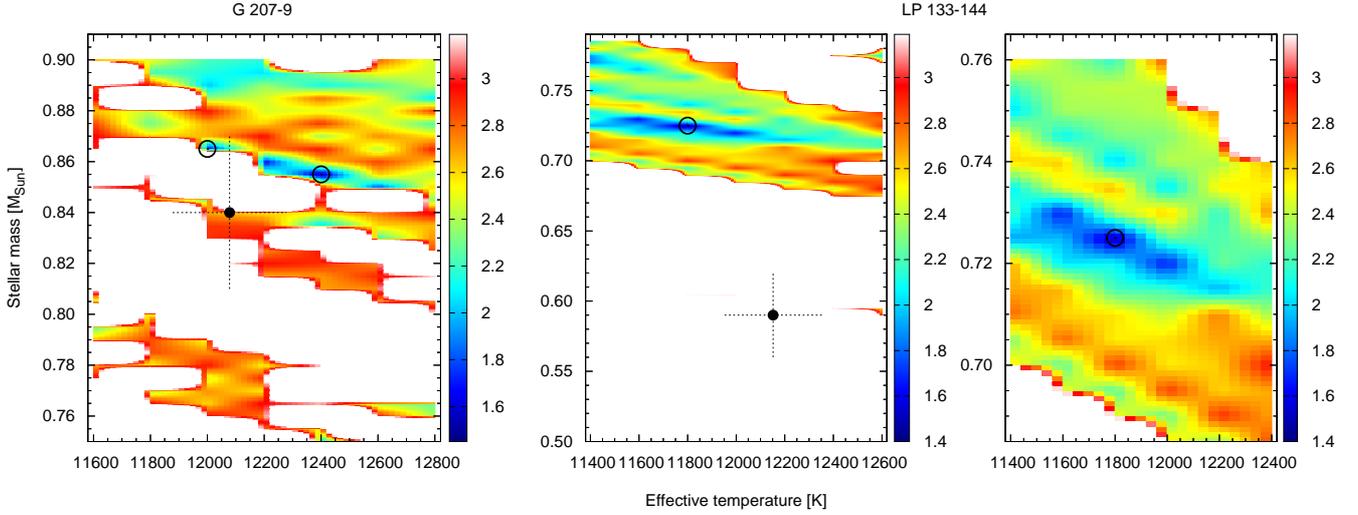}
\caption{Models on the $T_{\rmn{eff}}-M_*$ plane with their $\sigma_\mathrm{{rms}}$ values (colour coded). \textit{Left panel}: G~207-9; models 
fitting with eight periods and assuming that at least half of the modes is $l=1$, including the dominant mode. The two models with
the lowest $\sigma_\mathrm{{rms}}$ values (for more explanation, see the text) are denoted with open circles. Black dot marks
the spectroscopic value (cf. Table~\ref{table:g207params}). \textit{Middle panel}: LP~133-144; models fitting with six periods
and assuming that the three largest amplitude modes (also showing triplets) are $l=1$. The best-fitting model is denoted with
an open circle, the spectroscopic value presented in Table~\ref{table:lp133params} is signed with a black dot. \textit{Right panel}:
magnified part of the middle panel's plot around the best-matching model. We used interpolation in the plots for better visibility. 
}
\label{fig:grids}
\end{figure*}

In the case of four expected $l=1$ modes and seven periods, the best-matching model has the same effective temperature 
($T_{\rmn{eff}}=12\,000$\,K), a bit lower
stellar mass ($M_*=0.865\,M_{\sun}$), and thinner hydrogen layer ($M_\rmn{H}=10^{-6}\,M_*$).
Assuming four $l=1$ modes and eight periods, the best-matching model has $T_{\rmn{eff}}=12\,400$\,K, $M_*=0.855\,M_{\sun}$ 
and $M_\rmn{H}=10^{-4.6}\,M_*$. The second best-fit model is the same as for four $l=1$ modes and seven periods.
We denoted with open circles these two latter models in Fig.~\ref{fig:grids} (left panel) on the $T_{\rmn{eff}}-M_*$ plane, 
together with the spectroscopic solution.
Both in the case of G~207-9 and LP~133-144, we utilized the $T_{\rmn{eff}}$ and surface gravity ($\mathrm{log}\,g$) 
values provided by \citet{2011ApJ...743..138G}, and then corrected them
according to the results of \citet{2013AA...559A.104T} based on radiation-hydrodynamics 3D simulations of convective 
DA stellar atmospheres.
We accepted the resulting values as the best estimates for these atmospheric parameters.
We converted the surface gravities to stellar masses utilizing the theoretical masses determined
for DA stars by \citet{1996ApJ...468..350B}.

\begin{figure*}
\centering
\includegraphics[width=17.5cm]{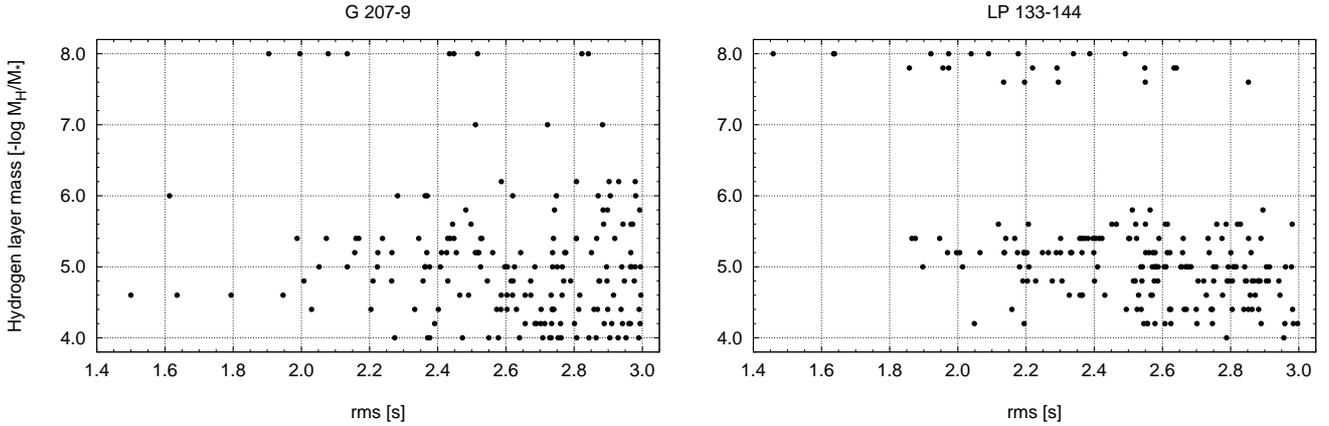}
\caption{Mass of the hydrogen layer of models at different $\sigma_\mathrm{{rms}}$ values.
\textit{Left panel}: G~207-9, \textit{right panel}: LP~133-144.
We plotted the same models that are presented in Fig.~\ref{fig:grids}.
}
\label{fig:grids2}
\end{figure*}

Considering the mass of the hydrogen layer (see the left panel of Fig.~\ref{fig:grids2}), we found that most of the models 
up to $\sigma_\mathrm{{rms}}=3.0$\,s
are in the $M_\rmn{H}=10^{-4}-10^{-6}\,M_*$ range, while about a dozen models predict thinner hydrogen layer down to $10^{-8}\,M_*$.
The best-fitting models favour the $M_\rmn{H}=10^{-4.6}\,M_*$ value.

We summarize the results of the spectroscopic atmospheric parameter determinations, the former modelling results based on 
the 1975 frequency list,
the main stellar parameters of the models mentioned above and the calculated periods fitted with our observed ones
in Table~\ref{table:g207params}. We also list the $\sigma_\mathrm{{rms}}$ values of the models.
The $T_{\rmn{eff}}=12\,000$\,K solutions are in agreement with the 
spectroscopic value.
The $T_{\rmn{eff}}=12\,400$\,K model seems somewhat too hot comparing to the $\sim12\,100$\,K spectroscopic
temperature, but considering that the uncertainties of both values are estimated to be around $200$\,K, 
this model still not contradicts to the observations.
The $0.855-0.870\,M_*$ stellar masses are also close to the value derived by spectroscopy, considering its uncertainty.
Summing it up, we can find models with stellar parameters and periods close to the observed values even if we 
assume that at least half of the modes is $l=1$, including the dominant mode.

\begin{table*}
\centering
\caption{G~207-9: results on the main stellar parameters obtained by spectroscopic observations (`\textit{Spectroscopy}'),
previous model fits (`\textit{Modelling}') and modelling presented in this paper (`\textit{This work}').
In the `\textit{Modelling}' part, we listed the observed periods used for the fits.
In the case of our model fits, we listed the observed periods first and present the theoretical periods at the actual
models. The $\sigma_\mathrm{{rms}}$ values are given at the effective temperatures in parentheses.
In the case of model fits the identification of pulsation modes - spherical degree ($l$) and radial order ($k$) values - 
are also presented.}
\label{table:g207params}
\begin{tabular}{llrrrr}
\hline
$T_{\rmn{eff}}$ (K) & \multicolumn{1}{c}{$M_*/M_{\sun}$} &  \multicolumn{1}{c}{-log\,$M_\rmn{He}$} & \multicolumn{1}{c}{-log\,$M_\rmn{H}$} & 
\multicolumn{1}{c}{Periods in seconds ($l$, $k$)} & Reference\\ 
\hline
\multicolumn{6}{l}{\textit{Spectroscopy:}}\\
12\,078$\pm$200 & 0.84$\pm$0.03 & & & & \citet{2011ApJ...743..138G}\\
 & & & & & \citet{2013AA...559A.104T}\\
\multicolumn{6}{l}{\textit{Modelling:}}\\
12\,000 & 0.815 & 2.0 & 8.5 & 259.0, 292.0, 317.3, 557.3, 740.7, 787.5$^\star$ & \citet{2009MNRAS.396.1709C}\\
11\,700 & 0.530 & 3.5 & 6.5 & 259.0, 292.0, 317.3, 557.3, 740.7, 787.5$^\star$ & \citet{2009MNRAS.396.1709C}\\
12\,030 & 0.837 & 2.5 & 6--7 & 259.1 (1,4), 292.0 (2,10), 318.0 (1,5),& \citet{2012MNRAS.420.1462R, 2013ApJ...779...58R}\\
 & & & & 557.4 (1,12), 740.4 (1,17) & \\
 & & & &  & \\
\multicolumn{2}{l}{\textit{This work:}} & & & \multicolumn{2}{l}{291.9, 595.7, 196.1, 623.8, 129.4, 317.8, 305.2, 557.4} \\
& & & &  & \\
12\,000 (1.06\,s) & 0.870 & 2.0 & 4.0 & 291.0 (1,7), 595.5 (2,32), 195.8 (2,9), & \\
& & & & 625.6 (2,34), 129.0 (2,5), 319.7 (2,16), & \\
& & & & 305.4 (2,13), 558.6 (2,28) & \\
12\,000 (1.61\,s) & 0.865 & 2.0 & 6.0 & 290.6 (1,5), 594.5 (1,14), 193.1 (2,6), & \\
& & & & 623.9 (1,15), 130.4 (2,3), 316.6 (1,6), & \\
& & & & 306.2 (2,12), 555.1 (2,24) & \\
12\,400 (1.50\,s) & 0.855 & 2.0 & 4.6 & 290.5 (1,6), 594.7 (1,16), 194.0 (1,3), & \\
& & & & 624.7 (1,17), 132.3 (2,4), 318.7 (2,14), & \\
& & & & 304.6 (2,13), 557.0 (2,27) & \\
\hline
\multicolumn{6}{l}{{\textbf{$^\star$}} The utilized periods were mean values of the periods of \citet{1976ApJ...207L..37R}
and the periods of WD~J0815+4437 showing}\\ 
\multicolumn{6}{l}{similar pulsation modes.}\\
\end{tabular}
\end{table*}

\subsection{Best-fitting models for LP~133-144}

The model with the lowest $\sigma_\mathrm{{rms}}$ ($0.46$\,s) has $T_{\rmn{eff}}=11\,800$\,K, $M_*=0.710\,M_{\sun}$ 
and $M_\rmn{H}=10^{-4.0}\,M_*$ if we do not apply any restrictions on the $l$ values of modes.
Generally, the best-matching models have masses around $0.7\,M_{\sun}$, which are at least $0.1\,M_{\sun}$
larger than the spectroscopic value. These models provide $3-4$ $l=1$ solutions to the 
observed modes.

We searched for the best-matching models in a second run, assuming that the three largest amplitude modes
showing triplet structures at 209.2, 305.9 and 270.6\,s are all $l=1$ modes. The best-matching model
has the same effective temperature ($T_{\rmn{eff}}=11\,800$\,K), slightly larger mass ($M_*=0.725\,M_{\sun}$) 
and much thinner
hydrogen layer ($M_\rmn{H}=10^{-8.0}\,M_*$) than the previously selected model. The mass still
seems too large comparing to the spectroscopic value, but it gives $l=1$ solutions for all the four modes
with triplet frequencies, including the mode at 179.5\,s. 
These modes are consecutive radial overtones with $k=1-4$.
We denoted this model with an open circle
on the middle and right panels of Fig.~\ref{fig:grids}.
The hydrogen layer masses versus the $\sigma_\mathrm{{rms}}$ values of these models are plotted in the right panel
of Fig.~\ref{fig:grids2}. This figure also shows that the best-fitting models have thin hydrogen layer with
$M_\rmn{H}=10^{-8.0}\,M_*$. Otherwise, two families of model solutions outlines: one with 
$M_\rmn{H}=10^{-4.0}-10^{-6.0}\,M_*$ and one with thinner, $M_\rmn{H}=10^{-7.6}-10^{-8.0}\,M_*$ hydrogen layers.

If we restrict our period fitting to the 
models with effective temperatures and masses being in the range determined by spectroscopy,
the best-matching model has $T_{\rmn{eff}}=12\,000$\,K, $M_*=0.605\,M_{\sun}$ 
and $M_\rmn{H}=10^{-4.2}\,M_*$. However, the 179.5\,s mode is $l=2$ in this case, while all the other
frequencies are consecutive radial order $l=1$ modes.

At last, we searched for models in this restricted parameter space and assuming that all the four frequencies 
showing triplets are $l=1$. Our finding with the lowest $\sigma_\mathrm{{rms}}$ has $T_{\rmn{eff}}=12\,000$\,K, 
$M_*=0.585\,M_{\sun}$ and $M_\rmn{H}=10^{-5.0}\,M_*$, however, its $\sigma_\mathrm{{rms}}$ is relatively 
large ($6.8$\,s), which means that there are major differences between the observed and calculated
periods. Table~\ref{table:lp133params} lists the stellar parameters and theoretical periods of the models 
mentioned above. For completeness, we included this last model solution, too.

We concluded, that our models predict at least $0.1\,M_{\sun}$ larger stellar mass for LP~133-144 than the
spectroscopic value. Nevertheless, it is possible to find models with lower stellar masses, but in these
cases not all the modes with triplet frequency structures has $l=1$ solutions and (or) the corresponding
$\sigma_\mathrm{{rms}}$ values are larger than for the larger mass models.
Considering the effective temperatures, the $T_{\rmn{eff}}=12\,000$\,K solutions are in agreement with
the spectroscopic determination ($\sim12\,150$\,K) within its margin of error. As in the case of G~207-9, 
taking into account that the uncertainties for the grid parameters are of the order of the step sizes in 
the grid, the $T_{\rmn{eff}}=11\,800$\,K findings are still acceptable.

\begin{table*}
\centering
\caption{Same as in Table~\ref{table:g207params} but for LP~133-144.}
\label{table:lp133params}
\begin{tabular}{llrrrr}
\hline
$T_{\rmn{eff}}$ (K) & \multicolumn{1}{c}{$M_*/M_{\sun}$} &  \multicolumn{1}{c}{-log\,$M_\rmn{He}$} & \multicolumn{1}{c}{-log\,$M_\rmn{H}$} & 
\multicolumn{1}{c}{Periods in seconds ($l$, $k$)} & Reference\\ 
\hline
\multicolumn{6}{l}{\textit{Spectroscopy:}}\\
12\,152$\pm$200 & 0.59$\pm$0.03 & & & & \citet{2011ApJ...743..138G}\\
 & & & & & \citet{2013AA...559A.104T}\\
\multicolumn{6}{l}{\textit{Modelling:}}\\
11\,700 & 0.520 & 2.0 & 5.0 & 209.2 (1,2), 305.7 (2,7), 327.3 (2,8) & \citet{2009MNRAS.396.1709C}\\
12\,210 & 0.609 & 1.6 & $\sim6$ & 209.2 (1,2), 305.7 (2,8), 327.3 (2,9) & \citet{2012MNRAS.420.1462R}\\
 & & & &  & \\
\multicolumn{2}{l}{\textit{This work:}} & & & \multicolumn{2}{l}{209.2, 305.9, 270.6, 327.3, 140.5, 179.5} \\
& & & &  & \\
11\,800 (0.46\,s) & 0.710 & 2.0 & 4.0 & 208.8 (1,3), 305.6 (2,11), 270.1 (1,5), & \\
& & & & 327.2 (1,6), 140.6 (2,4), 180.4 (1,2) & \\
11\,800 (1.46\,s) & 0.725 & 2.0 & 8.0 & 209.5 (1,2), 304.5 (1,4), 268.8 (1,3), & \\
& & & & 328.3 (2,9), 138.5 (2,2), 181.0 (1,1) & \\
12\,000 (2.89\,s) & 0.605 & 2.0 & 4.2 & 204.5 (1,2), 307.9 (1,4), 271.5 (1,3), & \\
& & & & 326.2 (1,5), 138.4 (1,1), 183.7 (2,4) & \\
12\,000 (6.83\,s) & 0.585 & 2.0 & 5.0 & 215.3 (1,2), 311.6 (1,4), 273.3 (1,3), & \\
& & & & 326.6 (2,9), 126.6 (2,2), 176.4 (1,1) & \\
\hline
\end{tabular}
\end{table*}

\subsubsection{Stellar rotation}
\label{sect:lp133rot}

A plausible explanation for the observed triplet structures is that these are rotationally split frequency components
of $l=1$ modes. We used this assumption previously in searching for model solutions for our observed periods.
Knowing the frequency differences of the triplet components ($\delta f$), we can estimate the rotation period of the pulsator.

In the case of slow rotation, the frequency differences of the $m=-1,0,1$ rotationally split components 
can be calculated (to first order) by the following relation: 

\begin{equation}
\label{eq:rot}
\delta f_{k,\ell,m} = \delta m (1-C_{k,\ell}) \Omega, 
\end{equation}
\noindent where the coefficient \mbox{$C_{k,\ell} \approx 1/\ell(\ell+1)$} for high-overtone ($k\gg\ell$) $g$-modes and 
$\Omega$ is the (uniform) rotation frequency.

In the case of LP~133-144, the presumed $l=1$ modes are low radial-order frequencies ($k=1-6$), but the $C_{k,\ell}$ values
of the fitted modes can be derived by the asteroseismic models. We used the average of the frequency separations within a triplet
and calculated the stellar rotation rate separately for $f_1$, $f_2$, $f_3$ and $f_6$ (see e.g. \citealt{2015MNRAS.451.1701H}).
We utilized the $T_{\rmn{eff}}=11\,800$\,K, $M_*=0.725\,M_{\sun}$ model.
The resulting rotation periods are: $P_{f_1} = 1.83$\,d ($\overline{\delta f_1}=4.15\,\mu$Hz, $C_{k,\ell}=0.345$),
$P_{f_2} = 1.82$\,d ($\overline{\delta f_2}=3.2\,\mu$Hz, $C_{k,\ell}=0.497$),
$P_{f_3} = 1.69$\,d ($\overline{\delta f_3}=3.5\,\mu$Hz, $C_{k,\ell}=0.489$) and 
$P_{f_6} = 1.60$\,d ($\overline{\delta f_6}=4.75\,\mu$Hz, $C_{k,\ell}=0.343$).
The average rotation period thus $1.74\pm0.11$\,d ($\sim42$\,h). This fits perfectly in the known rotation rates of 
the order of hours to days of ZZ Ceti stars (cf. Table~4 in \citealt{2008PASP..120.1043F}).
Note that the rotation periods calculated by the different multiplet structures are strongly depend on the 
actual values of observed frequency spacings and also on the $C_{k,\ell}$ values, which vary from model to model.
Thus the different rotation periods calculated for the different modes does not of necessarily mean that e.g. in this case 
we detected differential rotation of the star, but we can provide a reasonable estimation on the global rotation period
of LP~133-144.

\section{Summary and Conclusions}

We have presented the results of the one-season-long photometric observations of the ZZ Ceti stars G~207-9 and LP~133-144. These
rarely observed pulsators are located in the middle and in the hot part of the instability strip, respectively. 
G~207-9 was found to be a massive object previously by spectroscopic observations, comparing its predicted $M_*>0.8\,M_{\sun}$ 
mass to the average $\sim0.6\,M_{\sun}$ value of DA stars (see e.g. \citealt{2013ApJS..204....5K}).
In contrary, the mass of LP~133-144 was expected to be around this average value.

With our observations performed at Konkoly Observatory, we extended the number of known pulsation frequencies in both stars.
We found seven linearly independent modes in G~207-9, including five newly detected frequencies, comparing to
the literature data. 
We also detected the possible signs of additional frequencies around some of the G~207-9 modes, but their separations
being close to the 1\,d$^{-1}$ value makes their detection uncertain. Multi-site or space-based observations could verify
or disprove their presence.
In the case of LP~133-144, we detected three new normal modes out of the six derived, and revealed
that at least at three modes there are actually triplet frequencies with frequency separations of $\sim4\mu$Hz.

All the pulsation modes of LP~133-144 and most of the modes of G~207-9 are found to be below 330\,s, with amplitudes up 
to $\sim10$\,mmag. This fits to the well-known trend observed at ZZ Ceti stars that at higher effective temperatures
we see lower amplitude and shorter period light variations than closer to the red edge of the instability strip 
(see e.g. \citealt{2008PASP..120.1043F}). We also found that on the five-month time scale of our observations there
were no significant amplitude variations in either stars. This suits to their location in the instability domain again,
as short time scale large amplitude variations are characteristics of ZZ Cetis with lower effective temperatures. However, in the
case of G~207-9, the different frequency content of the 1975 and 2007 observations shows that amplitude variations 
do occur on decade-long time scale.

In addition, similar pulsational feature of the two stars is that both show light variations with one dominant mode ($A=10-11$\,mmag)
and several lower amplitude frequencies. 

The extended list of known modes allowed to perform new asteroseismic fits for both objects, in which we compared the observed
and calculated periods both with and without any restrictions on the $l$ values of modes. The best-matching models of G~207-9
have found to be close to the spectroscopic effective temperature and stellar mass, predicting $T_{\rmn{eff}}=12\,000$ 
or $12\,400$\,K and $M_*=0.855-0.870\,M_{\sun}$. For LP~133-144, the best-fitting models
prefer more than $0.1\,M_{\sun}$ larger stellar masses than the spectroscopic measurements
and $T_{\rmn{eff}}=11\,800$\,K effective temperatures. 
The main sources of the differences in our model solutions and the models 
presented by \citet{2009MNRAS.396.1709C}, even though they also used the \textsc{wdec}, 
can arise from the different periods utilized for the fits, the different core 
composition profiles applied, and the different way they determined the 
best-fitting models utilizing the amplitudes of observed periods as weights 
to define the goodness of the fits.
At last, we derived the
rotational period of LP~133-144 based on the observed triplets and obtained $P_{\rmn{rot}}\simeq42$\,h. 

Note that the results of the asteroseismic fits
presented in this manuscript are preliminary findings, and both
objects deserve more detailed seismic investigations utilizing the
extended period lists, similarly to the modelling presented 
for other hot DAV stars, GD 165 and Ross 548 \citep{2016ApJS..223...10G}. 
In the case of these objects, the authors
could identify models reproducing the observed periods quite well  
while staying close to the spectroscopic stellar parameters, and
also verified the credibility of the selected models in many other ways,
including the investigation of rotationally split frequencies.

\section*{Acknowledgements}

The authors thank the anonymous referee for the constructive comments on the manuscript.
The authors thank Agn\`es Bischoff-Kim for providing her version of the \textsc{wdec} and the \textsc{fitper} program.
The authors also thank the contribution of E. Bokor, \'A. Gy\H orffy, Gy. Kerekes, A. M\'ar and N. Sztan\-k\'o to the observations
of the stars.
The financial support of the Hungarian National Research, Development and Innovation Office (NKFIH) grants 
K-115709 and PD-116175, and the LP2014-17 Program of the Hungarian Academy of Sciences are acknowledged.
P.I.P. is a Postdoctoral Fellow of the The Research Foundation -- Flanders (FWO), Belgium.
L.M. and \'A.S. was supported by the J\'anos Bolyai Research Scholarship of the Hungarian 
Academy of Sciences.




\bibliographystyle{mnras}
\bibliography{g207_lp133_mnras} 




\appendix
\section{}
\label{app:g207}

Normalized differential light curves of G~207-9 obtained in 2007 at Piszk\'estet\H o mountain station of
Konkoly Observatory.

\section{}
\label{app:lp133}

Normalized differential light curves of LP~133-144 obtained in 2007 at Piszk\'estet\H o mountain station of
Konkoly Observatory.


\bsp	
\label{lastpage}
\end{document}